%% file: main.tex
\RequirePackage{fix-cm}
\documentclass[twocolumn]{svjour3}       % onecolumn (second format)
\smartqed  % flush right qed marks, e.g. at end of proof
\usepackage{graphicx}
\usepackage{newtxmath}
\usepackage{appendix}
\usepackage{nicefrac}

\makeatletter
\if@twocolumn
  \renewcommand\normalsize{%
   \@setfontsize\normalsize\@xpt{12.5pt}%
   \abovedisplayskip=3 mm plus6pt minus 4pt
   \belowdisplayskip=3 mm plus6pt minus 4pt
   \abovedisplayshortskip=0.0 mm plus6pt
   \belowdisplayshortskip=2 mm plus4pt minus 4pt
   \let\@listi\@listI}%

  \renewcommand\small{%
   \@setfontsize\small{8.5pt}\@xpt
   \abovedisplayskip 8.5\p@ \@plus3\p@ \@minus4\p@
   \abovedisplayshortskip \z@ \@plus2\p@
   \belowdisplayshortskip 4\p@ \@plus2\p@ \@minus2\p@
   \def\@listi{\leftmargin\leftmargini
               \parsep 0\p@ \@plus1\p@ \@minus\p@
               \topsep 4\p@ \@plus2\p@ \@minus4\p@
               \itemsep0\p@}%
   \belowdisplayskip \abovedisplayskip}

\else
  \if@smallext
   \renewcommand\normalsize{%
   \@setfontsize\normalsize\@xpt\@xiipt
   \abovedisplayskip=3 mm plus6pt minus 4pt
   \belowdisplayskip=3 mm plus6pt minus 4pt
   \abovedisplayshortskip=0.0 mm plus6pt
   \belowdisplayshortskip=2 mm plus4pt minus 4pt
   \let\@listi\@listI}%

  \renewcommand\small{%
   \@setfontsize\small\@viiipt{9.5pt}%
   \abovedisplayskip 8.5\p@ \@plus3\p@ \@minus4\p@
   \abovedisplayshortskip \z@ \@plus2\p@
   \belowdisplayshortskip 4\p@ \@plus2\p@ \@minus2\p@
   \def\@listi{\leftmargin\leftmargini
               \parsep 0\p@ \@plus1\p@ \@minus\p@
               \topsep 4\p@ \@plus2\p@ \@minus4\p@
               \itemsep0\p@}%
   \belowdisplayskip \abovedisplayskip}
 \else
  \renewcommand\normalsize{%
   \@setfontsize\normalsize{9.5pt}{11.5pt}%
   \abovedisplayskip=3 mm plus6pt minus 4pt
   \belowdisplayskip=3 mm plus6pt minus 4pt
   \abovedisplayshortskip=0.0 mm plus6pt
   \belowdisplayshortskip=2 mm plus4pt minus 4pt
   \let\@listi\@listI}%  

  \renewcommand\small{%
   \@setfontsize\small\@viiipt{9.25pt}%
   \abovedisplayskip 8.5\p@ \@plus3\p@ \@minus4\p@
   \abovedisplayshortskip \z@ \@plus2\p@
   \belowdisplayshortskip 4\p@ \@plus2\p@ \@minus2\p@
   \def\@listi{\leftmargin\leftmargini
               \parsep 0\p@ \@plus1\p@ \@minus\p@
               \topsep 4\p@ \@plus2\p@ \@minus4\p@
               \itemsep0\p@}%
   \belowdisplayskip \abovedisplayskip}
  \fi
\fi

\makeatother

\usepackage{microtype}
\begin{document}

\title{Medium Amplitude Parallel Superposition (MAPS) Rheology of a Wormlike Micellar Solution%\thanks{Grants or other notes
}


\author{Kyle R. Lennon        \and
        Gareth H. McKinley      \and
        James W. Swan%etc.
}


\institute{K.R. Lennon \at
              Department of Chemical Engineering \\
              Massachusetts Institute of Technology      %  \\
           \and
           G.H. McKinley \at
              Hatsopoulos Microfluids Laboratory \\
              Department of Mechanical Engineering \\
              Massachusetts Institute of Technology
           \and 
          J.W. Swan \at
              Department of Chemical Engineering \\
              Massachusetts Institute of Technology \\ 
              \email{jswan@mit.edu}
}

\date{Received: April 22, 2021}

\maketitle

\begin{abstract}
The weakly nonlinear rheology of a surfactant solution of wormlike micelles is investigated from both a modeling and experimental perspective using the framework of medium amplitude parallel superposition (MAPS) rheology. MAPS rheology defines material functions, such as the third order complex compliance, which span the entire weakly nonlinear response space of viscoelastic materials to simple shear deformations. Three-tone oscillatory shear deformations are applied to obtain feature-rich data characterizing the third order complex compliance with high data throughput. Here, data for a CPyCl solution are compared to the analytical solution for the MAPS response of a reptation-reaction constitutive model, which treats micelles as linear polymers that can break apart and recombine in solution. Regression of the data to the model predictions provides new insight into how these breakage and recombination processes are affected by shear, and demonstrates the importance of using information-rich data to infer precise estimates of model parameters.
\end{abstract}

\section{Introduction}
\label{intro}
The rheology of surfactant solutions of wormlike micelles (WLMs) has received considerable attention for decades, in part because of the commercial significance of these materials -- which can be found in many common household products such as shampoos and surface cleaners \cite{rubin-1983,smith-1995}, as well as being used in industrial applications such as oil extraction \cite{roberta-2015,yang-2002} -- and in part because certain aspects of their rheology are described by well-known phenomenological constitutive models, thus making them ideal model materials. Selecting and parameterizing an appropriate constitutive model for WLMs and comparing the response in a nonlinear flow enables us to tests the complete understanding of the microstructural physics of these materials and how this microstructure is connected to macroscopic mechanical properties.

While it has been shown that the linear viscoelastic response of multiple formulations of WLMs is consistent with a simple single-mode Maxwell model, multiple distinct constitutive models have been proposed to describe the nonlinear rheology of these materials. For instance, the Giesekus model has been employed to describe the medium amplitude oscillatory shear (MAOS) signatures \cite{gurnon-2012} and both the parallel (PS) and orthogonal superposition (OS) response of certain WLMs \cite{kim-2013}, while other authors have used the Oldroyd 8-constant framework to fit the nonlinear response of WLMs during the startup of steady shear flow \cite{saengow-2019}. The Oldroyd 8-constant framework can accurately describe the measured rheology because of the number of adjustable parameters \cite{oldroyd-1950}, but has no direct connection to the microstructural physics of WLMs. The Giesekus model, on the other hand, treats the elongated micelles in solution as a physical network of deformable flexible polymers \cite{giesekus-1982}. Though the stretching, reptation, and interactions between micelles can be understood in the same way as for polymers, WLMs in solution undergo additional distinct physical processes, such as breakage and recombination, which are not considered in the Giesekus model (hence the nomenclature `living polymers'). Thus both the Giesekus model and Oldroyd 8-constant framework are clearly deficient in their physical depiction of the wormlike microstructure.

Many of the deficiencies of the Giesekus and Oldroyd models are addressed in a constitutive model designed by Cates specifically for WLMs \cite{cates-1990}. This model again treats the micelles as reptating polymers in solution, with the key difference that it explicitly models the breakage and recombination reactions of micellar structures. This ``reptation-reaction'' (RR) model has been successfully applied to predict some properties of WLMs, such as the flow curve measured in steady shearing flow \cite{spenley-1993}. However, applications of the model are limited, in part due to it's formulation as an implicit integral equation and the complexity of the breakage and recombination survival functions, all of which render analytical and numerical solutions to the model more difficult than those for differential equations such as the Giesekus and Oldroyd models.

Despite the difficulty in solving the RR model for strongly nonlinear flows, it is possible to obtain exact analytical solutions to the model asymptotically. In this study, we will use the framework of medium amplitude parallel superposition (MAPS) rheology to obtain expressions for the first and third order response functions of the RR model that govern the weakly nonlinear behavior of the model in arbitrary simple shearing deformation protocols \cite{lennon-2020-1}. These response functions will be compared to experimental MAPS data obtained for a cetylpyridinium chloride solution using a recently-developed three-tone oscillatory deformation protocol. This three-tone protocol has been demonstrated to produce large and feature-rich data sets with high data throughput \cite{lennon-2020-2}, and a quantitative fit of the RR model to this expansive data set will be performed to gain new insight into the physics of the breakage and recombination processes in this WLM system.

\section{MAPS Rheology}

When a complex fluid with time-invariant properties is subjected to a simple shearing deformation, the resulting shear stress $\sigma(t)$ is a nonlinear functional of the imposed shear strain $\gamma(t)$. Just as analytic functions may be written as polynomials in the form of the Taylor series, the functional relationship between shear stress and the strain history possesses a polynomial expansion called the Volterra series. Within the framework of MAPS rheology, only the linear and leading order nonlinear terms in this expansion are considered, resulting in the following truncated frequency-domain Volterra series \cite{lennon-2020-1}:
\begin{align}
    & \hat{\sigma}(\omega) = G^*_1(\omega)\hat{\gamma}(\omega) \label{eq:MAPS} \\
    & \quad + \frac{1}{(2\pi)^2}\int\int\int_{-\infty}^{\infty}G^*_3(\omega_1,\omega_2,\omega_3)\hat{\gamma}(\omega_1)\hat{\gamma}(\omega_2)\hat{\gamma}(\omega_3) \nonumber \\
    & \quad\quad\quad\quad\quad\quad\quad \times \delta(\omega - \sum_{m=1}^3\omega_m)d\omega_1d\omega_2d\omega_3 + O(\hat{\gamma}^5). \nonumber
\end{align}
Here, we express the shear stress in terms of its Fourier transform:
\begin{equation}
    \hat{\sigma}(\omega) = \int_{-\infty}^{\infty}e^{-i\omega t}\sigma(t)dt,
\end{equation}
and likewise for the shear strain. The third-order truncated Volterra series reveals two response functions that govern the medium amplitude, simple shear response space. The first, $G^*_1(\omega)$, is the complex modulus familiar from linear viscoelastic theory. The response function $G^*_3(\omega_1,\omega_2,\omega_3)$ is called the third order complex modulus. Working in the frequency domain is convenient because it is often possible to directly obtain the analytical form for the third order complex modulus from asymptotic analysis of constitutive models, and because it is possible to directly measure this response function at discrete points in three-frequency space $(\omega_1,\omega_2,\omega_3)$ for real viscoelastic materials \cite{lennon-2020-2}. In the remainder of this work, we will demonstrate these salient features of MAPS rheology for the RR model and for a real surfactant solution of WLMs.

Though equation \ref{eq:MAPS} is written in terms of the complex moduli, there are other representations of MAPS rheology that may be more convenient in certain circumstances. These representations are obtained by instead writing the shear stress as a functional of the shear strain rate, or by writing the shear strain as a functional of the shear stress. We may define new response functions for each representation, but in fact all representations convey the same information. For instance, in the stress-controlled Volterra series obtained by swapping the stress and strain in equation \ref{eq:MAPS}, the third order response function is called the third order complex compliance, which is related directly to the linear and third order complex moduli:
\begin{align}
    & J^*_3(\omega_1,\omega_2,\omega_3) = \nonumber \\
    & \quad\quad -\frac{G^*_3(\omega_1,\omega_2,\omega_3)}{G^*_1(\omega_1)G^*_1(\omega_2)G^*_1(\omega_3)G^*_1(\omega_1+\omega_2+\omega_3)}.
    \label{eq:convert}
\end{align}
While in Section \ref{sec:solution} we find the solution for the third order complex modulus in the RR model, we employ the third order complex compliance in Section \ref{sec:results} to present stress-controlled MAPS data obtained experimentally for a solution of WLMs. For a more detailed description of the mathematical foundations of MAPS rheology, we refer readers to reference \cite{lennon-2020-1}.

\section{The Reptation-Reaction (RR) Model for Wormlike Micelles}

\subsection{Model Equations}
\label{sec:model}

In the RR model \cite{cates-1990}, the stress tensor $\boldsymbol{\sigma}$ obeys:
\begin{equation}
    \boldsymbol{\sigma} = \frac{15}{4}G_0\left[\boldsymbol{W} - \frac{1}{3}\boldsymbol{I}\right],
    \label{eq:stress}
\end{equation}
with the tensor $\boldsymbol{W}$ governed by the following integral equation:
\begin{equation}
    \boldsymbol{W} = \int_{-\infty}^{t} \mathcal{B}(v(t'))\exp\left[-\int_{t'}^{t}\mathcal{D}(v(t''))dt''\right]\boldsymbol{Q}(\boldsymbol{E}_{t't})dt'.
    \label{eq:model}
\end{equation}
The wormlike micellar structures are abstracted as segments of a tube encapsulating the micelle as it is oriented and deformed by the flow. New segments of tube may be created or destroyed by breakage or recombination processes, or by elongation of the micelles. The function $\mathcal{B}(v(t'))$ in equation \ref{eq:model} represents the rate at which tube segments were created at time $t'$ in the past, while the function $\mathcal{D}(v(t''))$ represents the rate at which tube segments were destroyed at time $t''$ in the past. The temporal dependence of these functions comes through the time-varying rate of tube retraction $v(t)$:
\begin{equation}
    v(t) = \boldsymbol{K}(t):\boldsymbol{W}(t),
    \label{eq:retraction}
\end{equation}
where $\boldsymbol{K}(t) = \nabla\boldsymbol{v}$ is the velocity gradient tensor. The tensor $\boldsymbol{Q}(\boldsymbol{E}_{t't})$ describes the effect of tube elongation, where $\boldsymbol{E}_{t't}$ is the deformation tensor describing the accumulated deformation between an arbitrary time $t'$ and the present time $t$. In this sense, equation \ref{eq:model} integrates the contributions to the stress tensor due to the elongation of tube segments created at time $t'$ with rate $\mathcal{B}(v(t'))$ over all past times, weighted by the cumulative tube segment survival probability described by an integral of $\mathcal{D}(v(t''))$.

While the functions $\mathcal{B}(v)$, $\mathcal{D}(v)$, and $\boldsymbol{Q}(\boldsymbol{E}_{t't})$ are prescribed by the RR model, they are quite complex. They are presented in full in Appendix \ref{app:model}. Asymptotic expansion of $\boldsymbol{Q}(\boldsymbol{E}_{t't})$ is analytically tractable, and will be used to obtain the solution for the third order complex modulus in the following section. However, it is not practical to use the full expressions for $\mathcal{B}(v)$ and $\mathcal{D}(v)$ in analytical and numerical studies of the model. Rather, these functions can be approximated with the constraint that in the absence of deformation, tube segments are created and destroyed at a rate set by the time scale $\tau$ associated with curvilinear diffusion of micelles within tube segments, $\mathcal{B}(0) = \mathcal{D}(0) = \tau^{-1}$, along with the constraints that $\mathcal{B}(v) \sim -v$ for large negative values of $v$, $\mathcal{D}(v) \sim v$ for large positive values of $v$, and:
\begin{equation}
    \mathcal{D} - \mathcal{B} = v,
    \label{eq:DB}
\end{equation}
which preserves the net rate of tube retraction.

Although this single integral model makes significant advances in modeling the physics of WLMs, it does nonetheless exclude certain physical processes, such as constraint release (CR) \cite{graham-2003}. Recently, a more detailed microscopic modeling framework for WLMs has been developed to include CR \cite{cates-2020}; however, this framework consists of a system of partial differential equations, for which analytical solutions are difficult to obtain even in the linear regime. Moreover, whether CR processes play an important role in the rheology of WLMs is not yet fully appreciated, and it is thought that CR may be suppressed or less important in these living polymer systems compared to in unbreaking entangled polymeric systems \cite{spenley-1993,milner-2001}. Therefore, in this study we neglect CR effects and consider only the single integral modeling framework of the original reptation-reaction model.

\subsection{MAPS Response}
\label{sec:solution}

Despite the complexity of the RR model compared to the simpler differential forms of the Giesekus and Oldroyd models, it is still possible to obtain an exact analytical solution for the third order complex modulus arising from the model. The detailed mathematical steps involved in obtaining this expression are given in the Supplementary Material, and the full expression for the solution is left to Appendix \ref{app:solution} due to its complexity. However, it is useful to note that the final expression for the third order complex modulus can be written as the sum of three contributions:
\begin{equation}
    G^*_3(\omega_1,\omega_2,\omega_3) = (\alpha - 1)G^*_{3,\mathcal{B}} + \alpha G^*_{3,\mathcal{D}} + G^*_{3,\boldsymbol{Q}},
    \label{eq:solution}
\end{equation}
where $G^*_{3,\mathcal{B}}$ results from the nonlinear dependence of the function $\mathcal{B}(v)$ on the deformation protocol, $G^*_{3,\mathcal{D}}$ from the nonlinear dependence of $\mathcal{D}(v)$ on the deformation protocol, and $G^*_{3,\boldsymbol{Q}}$ from the nonlinear dependence of the $\boldsymbol{Q}(\boldsymbol{E}_{t't})$ tensor on the deformation protocol. The parameter $\alpha$ is defined as follows:
\begin{equation}
    \alpha \equiv \left.\frac{d\mathcal{D}}{dv}\right\rvert_{v = 0} = 1 + \left.\frac{d\mathcal{B}}{dv}\right\rvert_{v = 0},
    \label{eq:RR_alpha}
\end{equation}
to be consistent with equation \ref{eq:DB}. Because the approximations for the creation and destruction functions $\mathcal{B}(v)$ and $\mathcal{D}(v)$ can take any form subject to the previously described constraints, $\alpha$ may be treated as the single adjustable parameter in the MAPS response of the RR model. The only other model parameters, $G_0$ and $\tau$, govern the linear response of the model, thus $\alpha$ is the only adjustable parameter that may be used to fit the model to weakly nonlinear data.

Equations \ref{eq:solution} and \ref{eq:RR_alpha} along with the expressions in Appendix \ref{app:solution} are significant in that they represent exact analytical solutions for the RR model in unsteady, weakly nonlinear shear flows. To the authors' knowledge, no other solutions for this model in unsteady nonlinear flows presently exist. It is especially significant that, due to the generality of MAPS rheology, the solution presented in this section is actually valid for all unsteady, weakly nonlinear shear flows of the RR model.

\section{Materials and Methods}

\subsection{Wormlike Micellar Solution}

The surfactant solution of wormlike micelles used in this study consists of cetylpyridinium chloride (CPyCl), sodium salicylate (NaSal), and sodium chloride (NaCl) in de-ionized water at concentrations of 100:60:33 mM. CPyCl and NaSal were supplied by Alfa Aesar, and reagent grade NaCl was purchased from Sigma Aldrich.

\subsection{Rheometry}

We perform a small amplitude oscillatory shear (SAOS) frequency sweep and three-tone MAPS tests on the WLMs using a DHR-3 Discovery Hybrid Rheometer from TA instruments, with TRIOS software v5.0.0. All tests were performed using a 60mm, 2$^{\circ}$ aluminum cone with a truncation gap of 58$\mu$m, with the lower Peltier plate maintained at 25$^{\circ}$C. To reduce the effects of solvent evaporation, the cone-and-plate fixture was covered by a solvent trap and sealed using hexadecane oil.

\subsection{Three-tone MAPS Experiments}
\label{sec:three_tones}

\begin{figure*}
    \centering
    \includegraphics[width=\textwidth]{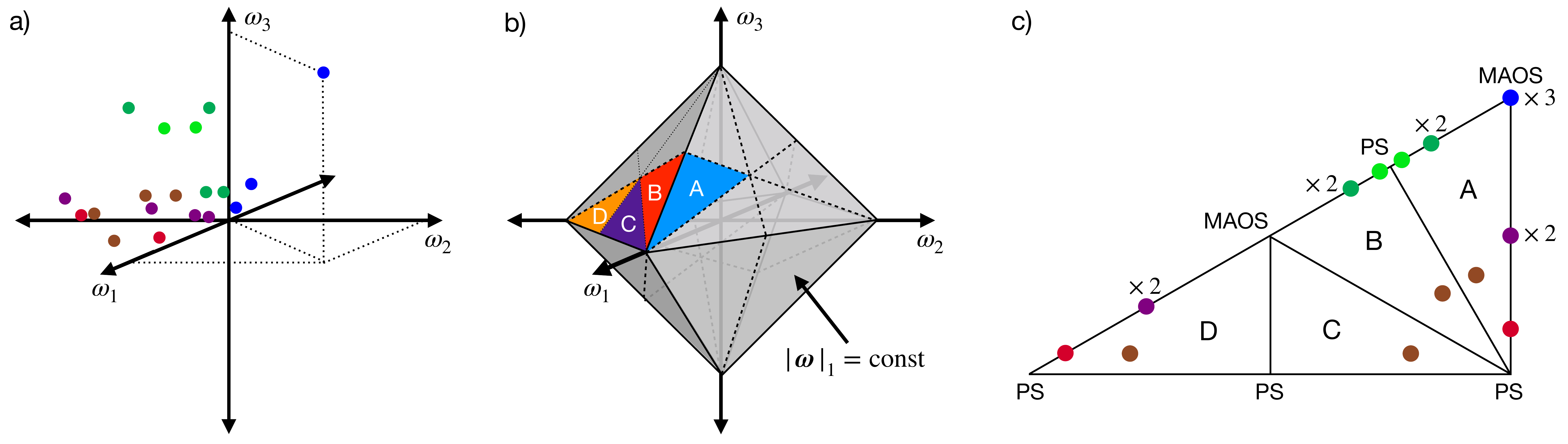}
    \caption{a) The distribution of points at which the third order complex compliance is measured by a three-tone MAPS experiment with input tones $\{n_1,n_2,n_3\} = \{1,4,16\}$. The selected value of the fundamental frequency $\omega_0$ of the three-tone input signal can rescale this distribution, but does not affect the relative distribution of measured points. b) A constant $L^1$-norm surface in three-frequency space. MAPS response functions possess symmetry with respect to permutation of their arguments, and Hermitian symmetry with respect to negation of their arguments. These symmetries reduce the domain on the $L^1$ surface on which MAPS response functions take unique values to the four colored subspaces: A, B, C, and D. c) Recording the positions at which each point in (a) intersects an $L^1$ surface as shown in (b) produces a 2D projection of coordinates that is independent of the imposed frequency scale. These projections are set uniquely by the integer triplet $\{n_1,n_2,n_3\}$. Within each subspace in this projection, we define a barycentric coordinate system which can be used to assign a unique RGB color to points within that subspace. These colors are shown in (c) together with the location of other weakly nonlinear deformation protocols such as medium amplitude oscillatory shear (MAOS) and parallel superposition (PS). For more details, see \cite{lennon-2020-2}.}
    \label{fig:maps}
\end{figure*}

To measure the third order complex compliance, we use the stress-controlled three-tone oscillatory shear experimental protocol developed in reference \cite{lennon-2020-2}. This experimental protocol is based on stress input signals of the form:
\begin{equation}
    \sigma(t) = \sigma_0 \left[\sin(n_1 \omega_0) + \sin(n_2 \omega_0) + \sin(n_3 \omega_0)\right].
\end{equation}
With the appropriate choice of the input parameters, these three-tone input signals yield 19 measurements of the third order complex compliance, $J^*_3(\omega_1,\omega_2,\omega_3)$, at distinct coordinates in three-frequency space.

The stress amplitude $\sigma_0$ may be selected based on amplitude sweep data to minimize the combined effect of bias from higher-order nonlinear effects and variance from instrumental noise. The fundamental frequency $\omega_0$ selects the time scale probed by the three-tone experiment, and the triplet of integers $\{n_1,n_2,n_3\}$ select how the 19 measurements are distributed throughout three-frequency space (independent of the frequency scale set by the chosen value of $\omega_0$). For example, the integer triplets $\{1,4,16\}$ produce the spread of points depicted in Figure \ref{fig:maps}a.

The measurements of $J^*_3(\omega_1,\omega_2,\omega_3)$ produced by these three-tone MAPS experiments are complex-valued and reside in a three-dimensional domain. Such high dimensional data is not straightforward to visualize. However, the data sets originating from MAPS experiments with a single integer set $\{n_1,n_2,n_3\}$ have a particular structure which allows the data to be plotted using familiar methods. This structure can be seen by examining where each measured point intersects a constant $L^1$-norm surface. A surface with a constant $L^1$-norm in three dimensions is equivalent to a regular octahedron as depicted in Figure \ref{fig:maps}b. Due to certain symmetries of the third order complex compliance, we can guarantee that all measured points will intersect this surface only in one of the four colored hemiequilateral triangles labeled A, B, C, and D \cite{lennon-2020-1}. For the example integer set $\{1,4,16\}$, recording the locations on a single projection of these four subspaces where data points intersect an $L^1$-norm surface, we obtain the projection of measured points shown in Figure \ref{fig:maps}c. This projection is determined uniquely by the selected integer set $\{n_1,n_2,n_3\}$ and is independent of the fundamental frequency $\omega_0$.

Using the depiction of data points in Figure \ref{fig:maps}, we may identify an alternative set of coordinates describing the location of each data point within three-frequency space. Within the projection of the four subspaces, we may record the subspace name $\mathcal{S}$ containing a given point, and within that subspace we may define a barycentric coordinate system with three barycentric coordinates -- $(r,g,b)$ -- specifying the location of the data point. The coordinate system is completed by recording the $L^1$-norm surface intersected by the data point, $|\boldsymbol{\omega}|_1 = |\omega_1| + |\omega_2| + |\omega_3|$. To visualize the data from a three-tone MAPS experiment, we can then create separate Bode or Nyquist plots for each subspace $\mathcal{S}$, and plot the magnitude and phase of the third order complex compliance as a function of $|\boldsymbol{\omega}|_1$. Points at different barycentric coordinates can be distinguished by color, with the coordinates $(r,g,b)$ specifying the RGB color of the points. Thus, by varying the fundamental frequency $\omega_0$ with constant integer set $\{n_1,n_2,n_3\}$, we impose frequency sweeps at each barycentric coordinate, which can be displayed on a shared set of axes. For more details on the experimental design and visualization of three-tone MAPS experiments, we refer readers to reference \cite{lennon-2020-2}.

\section{Results}
\label{sec:results}

\begin{figure}
    \centering
    \includegraphics[width=\columnwidth]{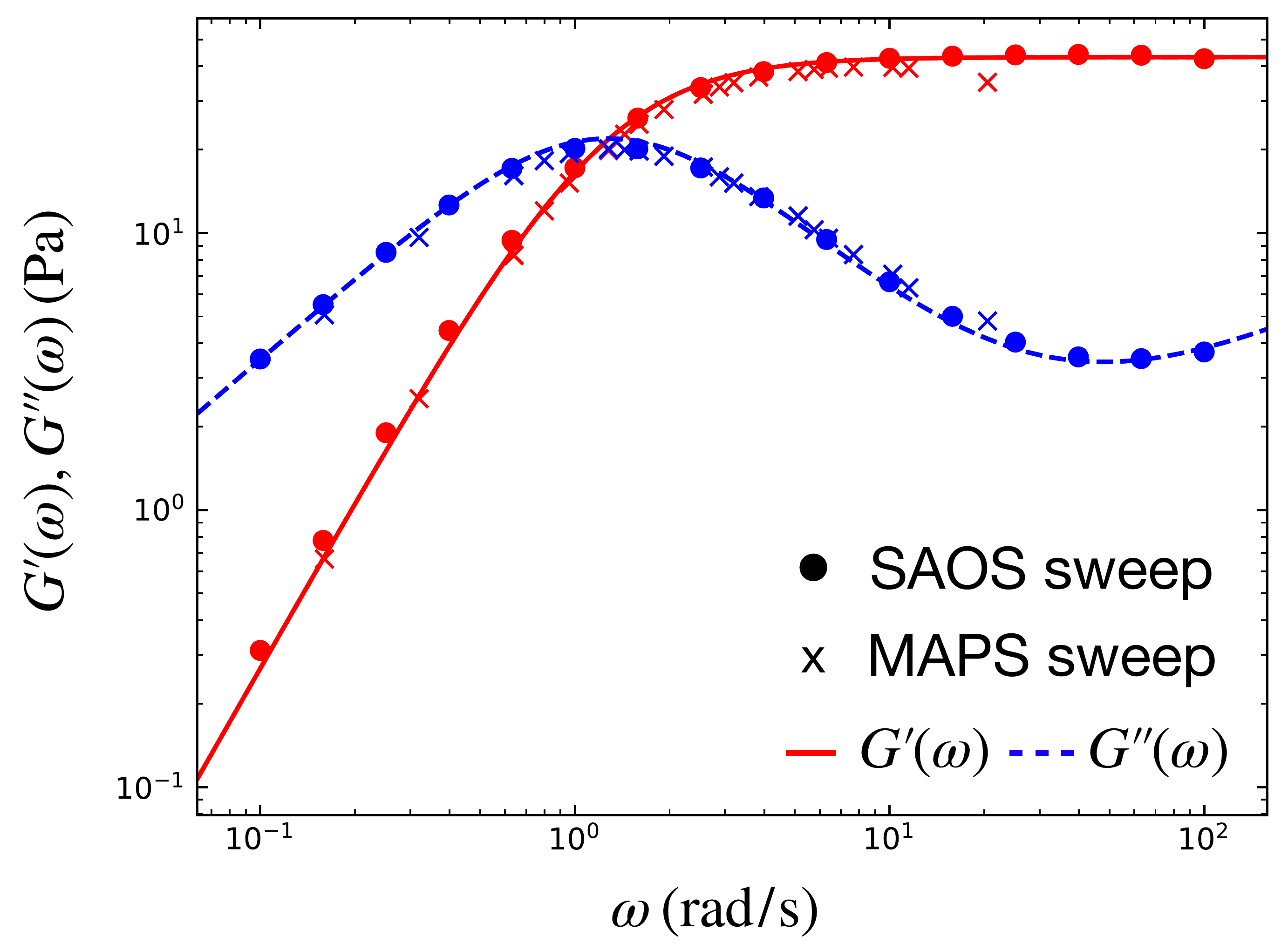}
    \caption{The linear storage modulus, $G'(\omega)$, and loss modulus $G''(\omega)$, of the CPyCl solution. Circles represent data obtained by a SAOS frequency sweep, and crosses represent data extracted from the three-tone MAPS experiments, with the storage modulus shown in red and loss modulus shown in blue. Data are compared to the predictions of the RR model with an added Rouse contribution, represented by solid red and dashed blue lines, respectively.}
    \label{fig:saos}
\end{figure}

The linear viscoelasticity of the wormlike surfactant solution was characterized via a stress-controlled small-amplitude oscillatory shear (SAOS) frequency sweep over the frequency range of 0.1 through 100 rad/s at a stress amplitude of 0.1 Pa. The measured linear storage and loss moduli, $G'(\omega)$ and $G''(\omega)$, are shown in Figure \ref{fig:saos}. The data are compared to the linear response of the RR model, which is equivalent to that of a single-mode Maxwell model. To describe the high-frequency upturn in the loss modulus, we superimpose a short-time Rouse contribution onto the predictions of the RR model \cite{cates-1990,doi-1986}. Thus, the storage and loss moduli are fit to the expressions:
\begin{subequations}
\begin{equation}
    G'(\omega) = \frac{G_0 \tau^2 \omega^2}{1 + \tau^2\omega^2},
\end{equation}
\begin{equation}
    G''(\omega) - c\omega^{\nicefrac{1}{2}} = \frac{G_0 \tau \omega}{1 + \tau^2\omega^2},
\end{equation}
\label{eq:linear}
\end{subequations}
The data are fit to this expression using least-squares regression, resulting in best-fit parameters of $G_0 = 43.2 \pm 0.9$ Pa, $\tau = 0.79 \pm 0.02$ s, and $c = 0.330 \pm 0.007$ Pa$\cdot$s$^{\nicefrac{1}{2}}$. Uncertainties in the parameter estimates reflect local estimates from the curvature of the least-squares objective function in the neighborhood of the minimum. These uncertainty estimates are discussed in more detail in the Supplementary Material.

\begin{figure*}
    \centering
    \includegraphics[width=\textwidth]{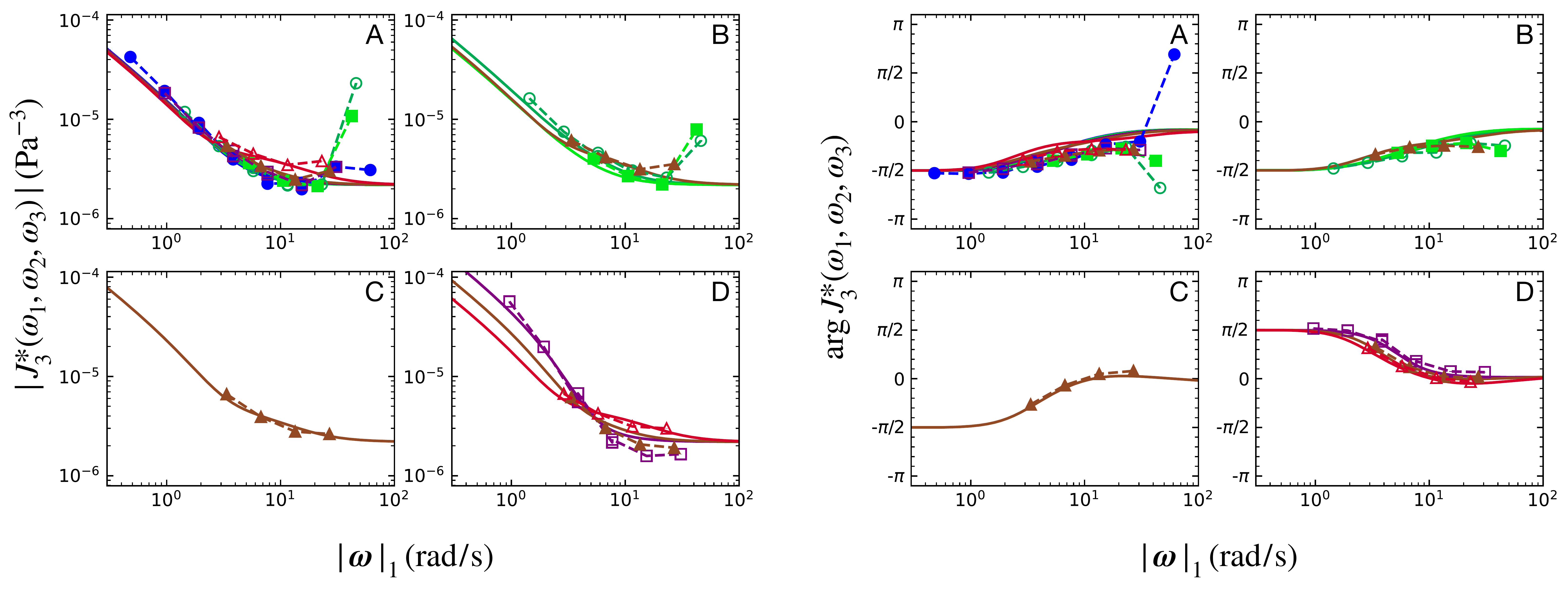}
    \caption{The magnitude (left) and phase angle (right) of the third order complex compliance measured for the wormlike micellar surfactant solution, obtained from a MAPS frequency sweep with $\{n_1,n_2,n_3\} = \{1,4,16\}$ over the fundamental frequencies $\omega_0 =$ 0.16, 0.32, 0.64, and 1.28 rad/s. The data are visualized according to the procedure described in Section \ref{sec:three_tones} and in \cite{lennon-2020-2}, with different colors also denoted by different symbols to aid in readability. Measured data are shown with symbols connected by dashed lines, and predictions of the RR model with the best fit value of $\alpha = -0.1$ are shown with solid lines.}
    \label{fig:1416}
\end{figure*}
\begin{figure*}
    \centering
    \includegraphics[width=\textwidth]{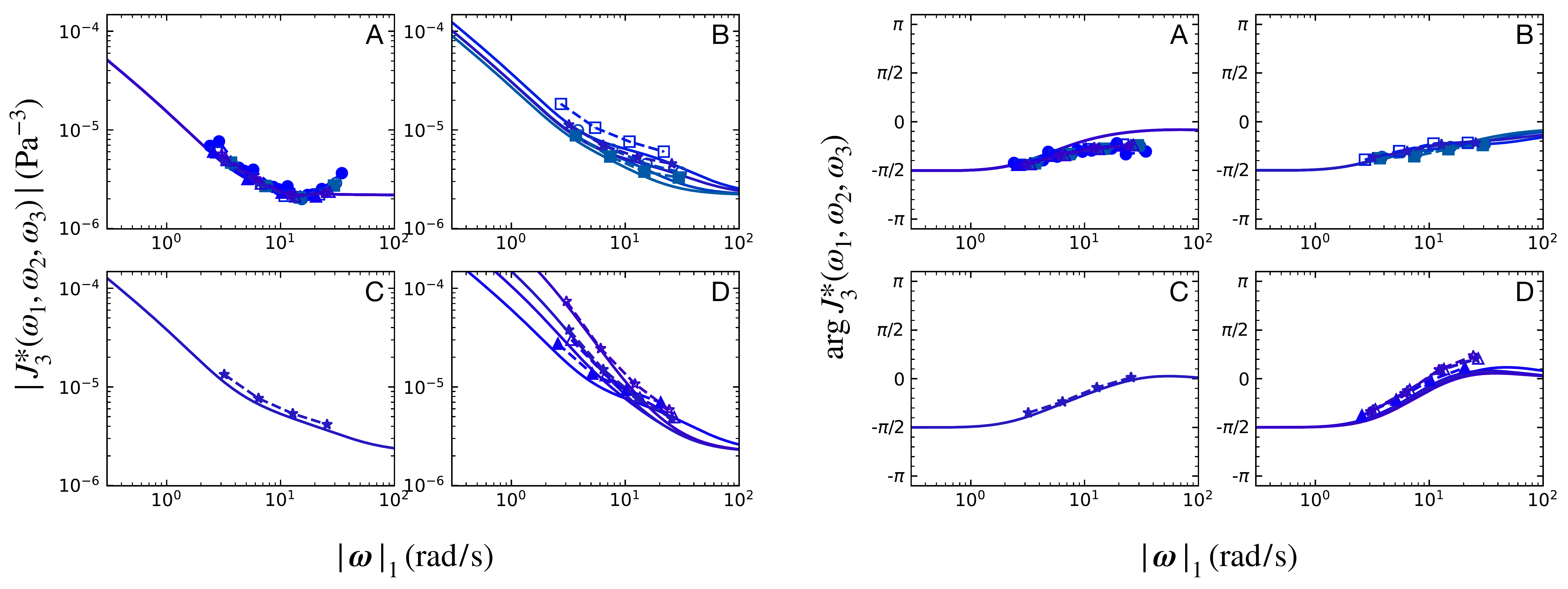}
    \caption{The magnitude (left) and phase angle (right) of the third order complex compliance measured for the wormlike micellar surfactant solution, obtained from a MAPS frequency sweep with $\{n_1,n_2,n_3\} = \{5,6,9\}$ over the fundamental frequencies $\omega_0 =$ 0.16, 0.32, 0.64, and 1.28 rad/s. The data are visualized according to the procedure described in Section \ref{sec:three_tones} and in \cite{lennon-2020-2}, with different colors also denoted by different symbols to aid in readability. Measured data are shown with symbols connected by dashed lines, and predictions of the RR model with the best fit value of $\alpha = -0.1$ are shown with solid lines.}
    \label{fig:569}
\end{figure*}

Even without the additional contribution of the Rouse modes at high frequency, the single-mode Maxwell predictions describe the LVE data remarkably well. This agreement has been noted by numerous other authors \cite{gurnon-2012,kim-2013,saengow-2019,vasquez-2007,pipe-2010}. That the RR model reduces to this prediction in the linear regime is encouraging, but alone does not distinguish it as more descriptive of the rheology of WLMs than the Giesekus or Oldroyd models, which also reduce to a single-mode Maxwell model in the linear regime. To assess whether the RR model is truly able to capture the rheological signatures of real WLMs requires measurements in the nonlinear regime.

Three-tone MAPS experiments, as described in Section \ref{sec:three_tones}, were conducted to study the nonlinear response of the WLM solution. To exemplify the richness of data from these tests, we run separate MAPS frequency sweeps with the integer triplets $\{1,4,16\}$ and $\{5,6,9\}$, over the fundamental frequencies $\omega_0 =$ 0.16, 0.32, 0.64, and 1.28 rad/s. These tests were each run at separate stress amplitudes of $\sigma_0 = 0.7$ Pa and two replicates at $\sigma_0 = 0.35$ Pa, and the linear and third order response regressed from these tests using the polynomial interpolation procedure described in \cite{lennon-2020-2}. The real and imaginary components of the linear viscoelastic complex modulus obtained from the three-tone MAPS experiments are presented along with the (separately measured) small-amplitude sweep data in Figure \ref{fig:saos}. The close agreement between these two independent data sets is a positive indication that a cubic polynomial sufficiently describes the weakly nonlinear data without substantial bias from higher-order nonlinearities. The third order complex compliance data are presented in Figures \ref{fig:1416} and \ref{fig:569} using the visualization scheme previously discussed.

The MAPS data are compared to predictions of the RR model, obtained by substituting the solution for the third order complex modulus in equation \ref{eq:solution}, plus the linear complex modulus in equation \ref{eq:linear}, into equation \ref{eq:convert}, using the parameters $G_0$, $\tau$, and $c$ obtained from the linear response. The single adjustable parameter $\alpha$ in the MAPS response function was used to fit the model predictions to the data using a weighted least-squares scheme, which has been previously suggested as an effective and unbiased estimator for rheological data fitting \cite{singh-2019}:
\begin{equation}
    \hat{\alpha} = \underset{\alpha}{\operatorname{arg\,min}}\left[(\hat{\textbf{J}}(\alpha) - \tilde{\textbf{J}})^{T}\boldsymbol{\Sigma}^{-1}(\hat{\textbf{J}}(\alpha) - \tilde{\textbf{J}})\right],
\end{equation}
where $\hat{\textbf{J}}(\alpha)$ is a vector obtained by vertically concatenating the real and imaginary parts of the model prediction for the third order complex compliance with a particular value of $\alpha$, $\tilde{\textbf{J}}$ is a vector obtained by vertically concatenating the real and imaginary parts of the experimental data set, and $\boldsymbol{\Sigma}$ is the covariance matrix of the data. Because the model prediction at third order is linear in $\alpha$, this global minimum can be found analytically, and is in this case found to be $\hat{\alpha} = -0.1 \pm 0.2$. With reference to equations \ref{eq:DB} and \ref{eq:RR_alpha}, this corresponds to a relatively small slope in the destruction function $\mathcal{D}(v)$ at low shear rates (when $v \rightarrow 0$), and a substantially steeper zero-shear-rate slope in the creation function $\mathcal{B}(v)$. The frequency-dependent trends for the third order complex compliance predicted by the RR model with $\alpha = -0.1$ are presented alongside the experimental MAPS data in Figures \ref{fig:1416} and \ref{fig:569}.

\begin{figure}
    \centering
    \includegraphics[width=\columnwidth]{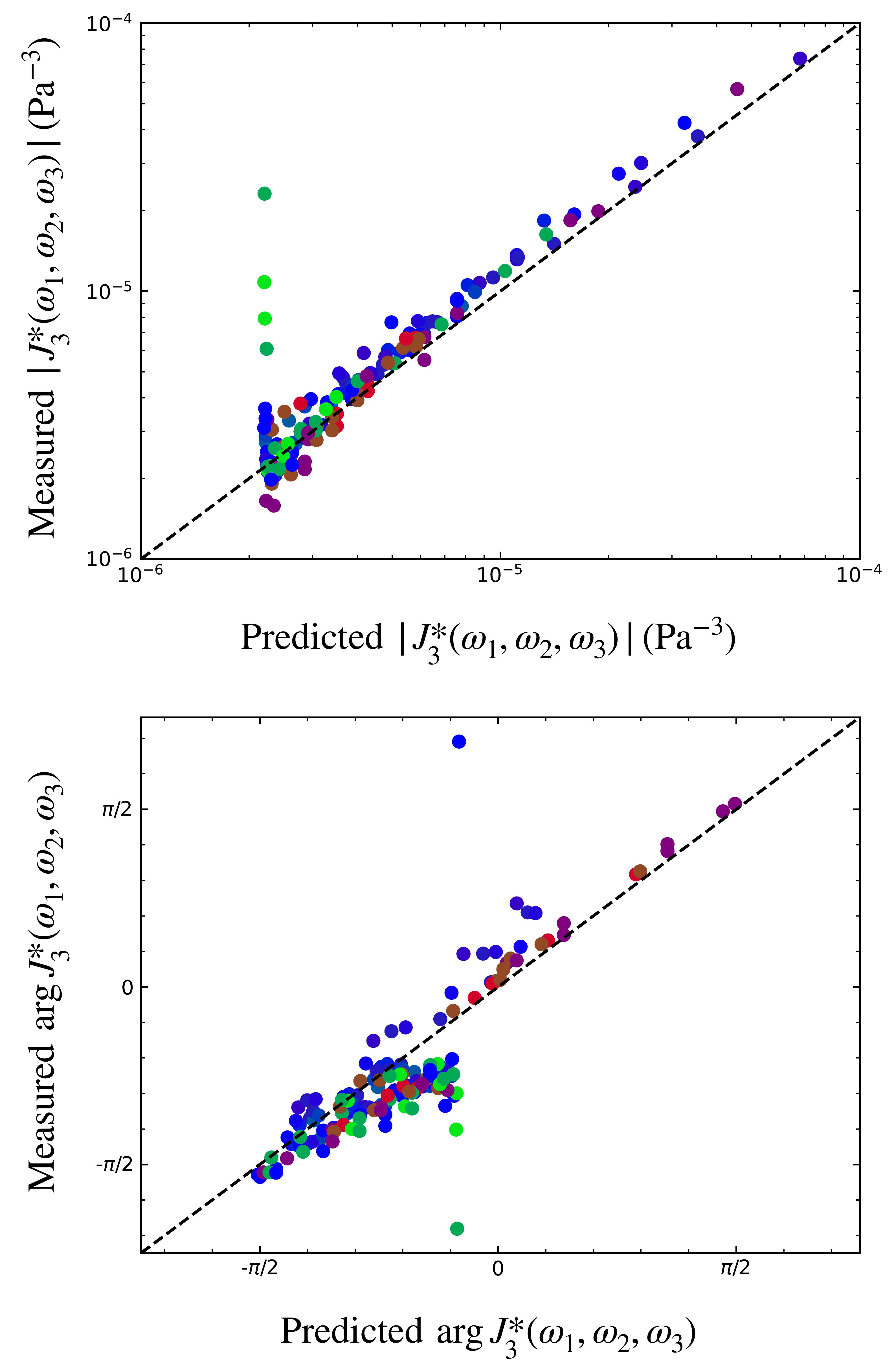}
    \caption{Parity plots of the magnitude (top) and phase angle (bottom) of the third order complex compliance, $J^*_3(\omega_1,\omega_2,\omega_3)$, comparing the values measured by the three-tone MAPS experimental protocol to the values predicted by the RR model with $\alpha = -0.1$. Data from both MAPS frequency sweeps (Figures \ref{fig:1416} and \ref{fig:569}) are shown on the same axes, with points colored according to the scheme depicted in Figure \ref{fig:maps}.}
    \label{fig:parity}
\end{figure}

The close agreement between the experimental data and the model predictions with $\alpha = -0.1$ highlights the strength of the RR model in describing the rheology of this WLM solution. In nearly every case, the model is accurate to within a factor of two in predicting the magnitude of the measured third order complex compliance, and is very often even more accurate than that. Moreover, the trends with respect to both varying the frequency $L^1$-norm and with respect to changing barycentric coordinates that are seen in the data are also predicted by the RR model. The same is true of the phase angle. The agreement between measurements and model predictions across the entire data set can be captured compactly by computing parity plots of the measured and predicted magnitude and phase angle of the third order complex compliance, as presented in Figure \ref{fig:parity}. In the parity plot for the measured and predicted magnitude, nearly all data points are tightly scattered around the line of parity. The only small systematic deviations occur when the model slightly under-predicts the data points with the largest magnitude, corresponding to measurements at lower $|\boldsymbol{\omega}|_1$. The data and model predictions for the phase angle also lie close to parity. Some slight systematic deviation from parity is observed, which by comparison with Figures \ref{fig:1416} and \ref{fig:569} can be associated with the data points measured at the highest imposed frequency. The few data points that deviate strongly from parity -- namely four of the green colored points in the parity plot of the magnitude, and two green and one blue data points in the parity plot of the phase angle -- may not reflect a deficiency in the model, but rather instrumental limitations, such as bias from the moment of inertia of the cone-and-plate fixture, or approaching the gap loading limit for high-frequency nonlinear measurements \cite{lennon-2020-2}. Still, the observation that such close agreement between the RR model and MAPS data can be achieved by determining only a single adjustable parameter is striking. Moreover, this observation, combined with the observation that the linear rheology in Figure \ref{fig:saos} is well-described by a single-mode Maxwell model, is consistent with the suggestion that constraint release is indeed suppressed in this system.

\begin{figure}
    \centering
    \includegraphics[width=\columnwidth]{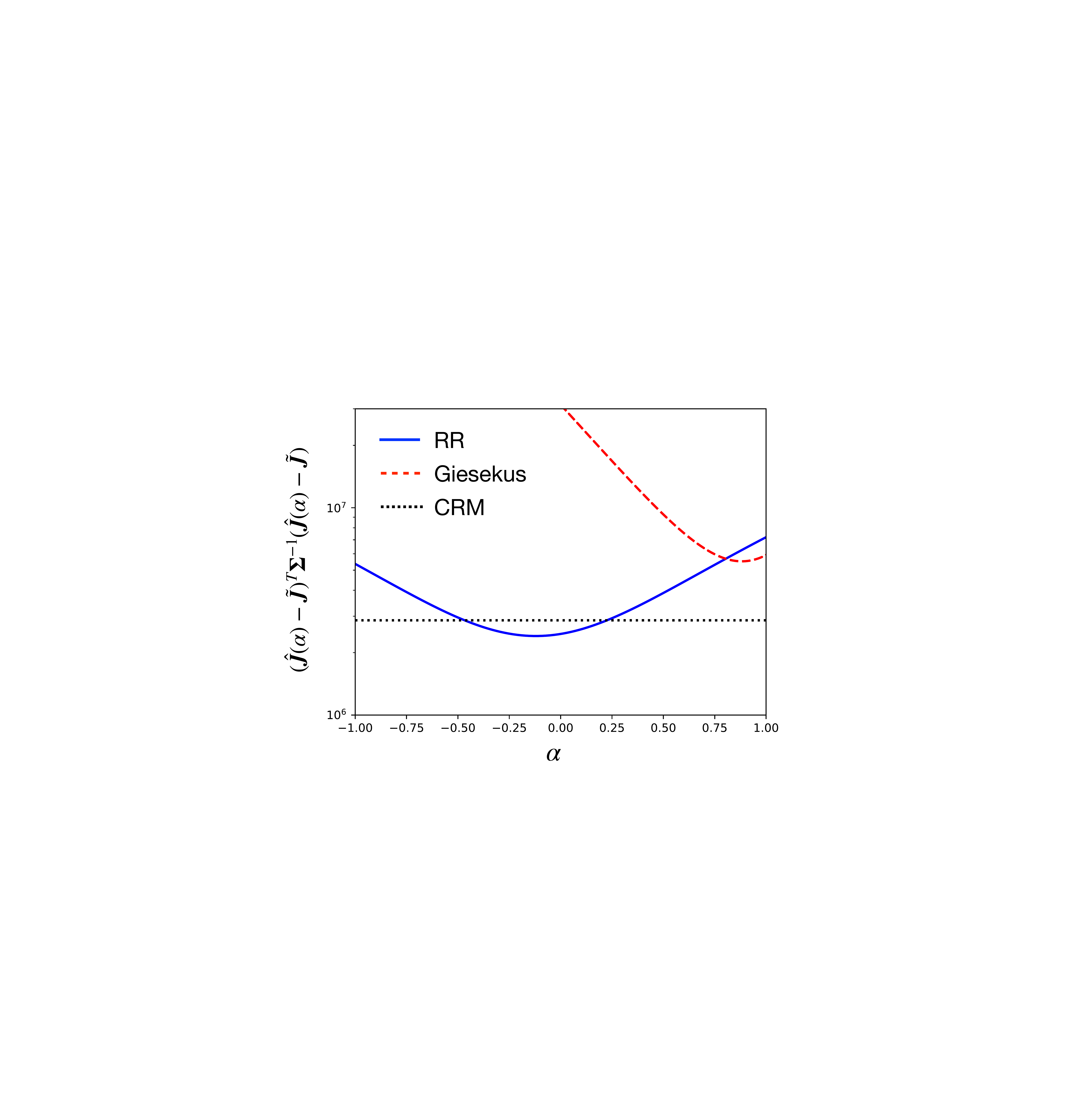}
    \caption{The weighted sum-of-squares error between the model predictions $\hat{\textbf{J}}(\alpha)$ and the experimental MAPS data $\tilde{\textbf{J}}$ as a function of the single adjustable parameter $\alpha$ in the RR and Giesekus models. The corresponding value of the weighted sum-of-squares error for the corotational Maxwell model (CRM), which does not have any adjustable parameters to fit the MAPS data, is shown with a horizontal dashed line.}
    \label{fig:error}
\end{figure}

The fit of the RR model to the experimental data is even more impressive when we also consider how the data compares to the predictions of other common constitutive models. The Giesekus model has been previously shown to predict a frequency dependence in both the magnitude and phase angle of the third order complex compliance that is inconsistent with the data shown in Figures \ref{fig:1416} and \ref{fig:569} \cite{lennon-2020-2}. This observation is further affirmed by examining parity plots of the Giesekus model prediction, akin to those in Figure \ref{fig:parity}, which are presented in the Supplementary Material and show the systematic deviations between observation and prediction. The corotational Maxwell model provides a better qualitative description of the experimental MAPS data, although the results are still not as good a quantitative description as the best fit to the RR model. This result can be rationalized by observing that the third order complex modulus previously derived for the corotational Maxwell model \cite{lennon-2020-1} closely resembles the contribution $G^*_{3,\boldsymbol{Q}}$ to the third order complex modulus of the RR model (equation \ref{eq:solution}), differing only by a constant factor of $7/6$ (as derived in the Supplementary Material). This slight difference in magnitude, combined with enhancements in the predictions from the terms $G^*_{3,\mathcal{B}}$ and $G^*_{3,\mathcal{D}}$, account for the closer fit and lower residual error of the RR model to the MAPS data. Other models in the Oldroyd 8-constant framework may also be able to reduce the discrepancy between the corotational Maxwell model predictions and experimental data, but such fits lack the microstructural physical basis that is gained from the RR model. 

To demonstrate the superiority of the RR model in describing the data quantitatively, we plot the value of the weighted sum-of-squares objective as a function of the adjustable parameter $\alpha$ in the RR and Giesekus models in Figure \ref{fig:error}. For the RR model, we display predictions in the range $\alpha \in [-1,1]$, and for the Giesekus model we restrict $\alpha \in [0,1]$ to maintain physical relevance. The weighted sum-of-squares error computed from the corotational Maxwell model, which has no adjustable parameters that can be used to improve the fit to the MAPS data, is shown with a dashed horizontal line. The smallest value of the weighted sum-of-squares error from the RR model is clearly less than the minimal error obtained from either the corotational Maxwell model or the Giesekus model, providing a quantitative confirmation that the RR model provides the best fit of the models considered here to the experimental data obtained using our MAPS protocol.

Because the RR model is derived from a microstructural model, we may interpret the optimal estimate for $\alpha$ in terms of real physical processes. Specifically, $\alpha$ relates to the asymptotic dependence of the rates of creation and destruction of tube segments on the rate of tube contraction. In his presentation of the model, Cates assumes that the functions approximating the creation and destruction rates are symmetric about $v = 0$ -- that is, that a small increase to the net rate of tube retraction is equivalently a result of decreased tube creation rate and increased tube destruction rate. This symmetric case is described by $\alpha = 0.5$. However, our best estimate of $\alpha$ is slightly below zero, indicating that these creation and destruction processes are not symmetric. From equation \ref{eq:RR_alpha}, the best fit value of $\alpha = -0.1 \pm 0.2$ actually indicates that small increases in the net retraction rate are mostly due to a decreased rate of tube segment creation as the shear rate is increased, and that the rate at which tube segments are destroyed does not initially change much as the material is deformed. This asymmetry may be due, for instance, to different free energy barriers required for the breakage and combination of micelles. Though a more detailed study of this phenomenon is outside the scope of this work, we note that insights such as this are not easy to obtain in other modelling frameworks, but are facilitated by the sensitivity of MAPS data to this specific feature of the model.

To reach this conclusion regarding asymmetric tube creation and destruction processes requires that a sufficiently sensitive estimate of $\alpha$ can be extracted from experimental data, and that the hypothesis of symmetric creation/destruction processes (corresponding to $\alpha = 0.5$) lies outside the window of uncertainty. Due to the large size and feature-rich nature of MAPS data sets, it was possible to make an estimate with this precision. The sensitivity is also evident in Figure \ref{fig:error}, which demonstrates that the weighted sum-of-squares error in the data set using the symmetric hypothesis of $\alpha = 0.5$ is nearly double the error of the best fit, and is actually greater than the error for the simpler corotational Maxwell model. The same sensitivity may not be present in other tests. For example, if we use only the 24 data points present on the third harmonic of each input tone in the three-tone protocol (corresponding to three points from each of the eight experiments with distinct $\omega_0$ and $\{n_1,n_2,n_3\}$) to estimate $\alpha$, reminiscent of the third harmonic data that would be available in medium amplitude oscillatory shear (MAOS) tests \cite{ewoldt-2013}, our estimate would instead be $\hat{\alpha} = -0.2 \pm 0.7$. The uncertainty in this estimate is now substantially higher than the case when all 152 MAPS data points are included (corresponding to 19 points from each of the eight combinations of $\omega_0$ and $\{n_1,n_2,n_3\}$), and now is too large to effectively rule out the symmetric case of $\alpha = 0.5$. This is despite the fact that obtaining the same number of third harmonic MAOS data points using single-tone tests would actually require more experiments than were used to obtain all of the MAPS data in Figures \ref{fig:1416} and \ref{fig:569}. Thus, three-tone MAPS tests are able to provide information that is more sensitive to specific nonlinear features of interest without requiring more experimental work than other weakly nonlinear techniques.

\section{Conclusions}

Constitutive modeling plays a critical role in understanding complex fluids and engineering their material characteristics. Selecting the model that is most appropriate for, and most descriptive of, a certain material is a difficult and sometimes ambiguous task, but nonetheless an important one. In many cases, it is desirable to choose a microstructural model originating from physical considerations about the material under study, as this provides a conduit to understanding specific structural characteristics of the material that may be designed to target specific material properties. However, for many constitutive models derived from microstructural physics, such as the RR model developed by Cates for wormlike micellar fluids, quantitative comparisons to data can be quite difficult due to their more complex mathematical structure compared to simpler, phenomenological models. Without analytical solutions to certain flows, for instance, tasks such as comparing the model response to measured data and particularly parameter estimation using experimental data become much more laborious.

Using asymptotic analysis and the MAPS rheology framework, however, it is often possible to obtain analytical solutions for complicated constitutive models at third order. In this work, we have used these tools to obtain analytical solutions for the RR model in a medium amplitude oscillatory flow for the first time. These asymptotic solutions enabled us to construct a weighted least-squares parameter estimation problem that could be solved exactly, and which provided detailed insight into a distinct physical feature driving the weakly nonlinear rheological signatures of WLMs. This insight was made possible by the large and feature-rich MAPS data set obtained using a three-tone deformation protocol. From constitutive modeling to experimentation, this study has demonstrated that the MAPS rheology framework can be a powerful tool for understanding and describing the nonlinear shear response of complex fluids.

\section*{Acknowledgements}

K.R.L. was supported by the U.S. Department of Energy Computational Science Graduate Fellowship program under Grant No. DE-SC0020347.

%\begin{acknowledgements}
%\end{acknowledgements}

\bibliographystyle{ieeetr}      % basic style, author-year citations
\bibliography{bibliography.bib}   % name your BibTeX data base

\input{appendix.tex}

\end{document}

%% file: appendix.tex
\appendix

\section{The Reptation-Reaction Model Equations}
\label{app:model}

The stress tensor in the reptation-reaction model is specified by equations \ref{eq:stress} and \ref{eq:model}, along with the companion equation \ref{eq:retraction}. Completing the model specifications requires equations for the tube creation rate $\mathcal{B}(v)$, destruction rate $\mathcal{D}(v)$, and the evolution of the $\boldsymbol{Q}(\boldsymbol{E}_{t't})$ tensor describing tube elongation. The creation and destruction rates are specified by two ensemble-averaged integrals over the position of the end of a micelle, $X(t)$:
\begin{equation}
    \mathcal{D} = \frac{2}{\bar{L}\hat{\tau}}\left\langle\int_0^{\infty}e^{-t/\hat{\tau}}\max_{0<t'<t}[X(t')]dt\right\rangle,
\end{equation}
\begin{equation}
    \mathcal{B} = \frac{2}{\bar{L}\hat{\tau}}\left\langle\int_0^{\infty}e^{-t/\hat{\tau}}\left(\max_{0<t'<t}[X(t')] - X(t)\right)dt\right\rangle.
\end{equation}
Angular brackets here represent ensemble averages over the stochastic processes of diffusion and reactions at the end of the micelles, such as breakage and recombination. In these expressions, $\bar{L}$ represents the time-averaged micelle length $L(t)$, and $\hat{\tau}$ a time scale related to the recombination process (this time scale is distinct from the time scale $\tau$ defined in Section \ref{sec:model}, which reflects a time scale associated with diffusion of the end of the micelle).

The tensor $\boldsymbol{Q}(\boldsymbol{E}_{t't})$ represents the average orientation over an isotropic distribution of unit vectors $\boldsymbol{u}$:
\begin{equation}
    \boldsymbol{Q}(\boldsymbol{E}_{t't}) = \frac{1}{4\pi}\int_{\mathcal{S}}\frac{[\boldsymbol{E}_{t't}\cdot\boldsymbol{u}][\boldsymbol{E}_{t't}\cdot\boldsymbol{u}]}{\lvert\boldsymbol{E}_{t't}\cdot\boldsymbol{u}\rvert}d^2\boldsymbol{u},
\end{equation}
where $\mathcal{S}$ represents the surface of the unit sphere, and $|\boldsymbol{x}|$ represents the $L^2$-norm of a vector $\boldsymbol{x}$.

\section{MAPS Response of the Reptation-Reaction Model}
\label{app:solution}

The third order complex modulus of the reptation-reaction model may be written as the sum of three terms:
\begin{equation}
    G^*_3(\omega_1,\omega_2,\omega_3) = (\alpha - 1)G^*_{3,\mathcal{B}} + \alpha G^*_{3,\mathcal{D}} + G^*_{3,\boldsymbol{Q}}.
\end{equation}
The term $G^*_{3,\mathcal{B}}$ represents the contribution due to nonlinearity in the tube creation function $\mathcal{B}(v)$:
\begin{align}
    & G^*_{3,\mathcal{B}}(\omega_1,\omega_2,\omega_3) = \\
    & \quad -\frac{2 G_0}{45}\sum_j \sum_{k\neq j}\frac{\omega_j\omega_k\tau^2}{1 + i\tau\omega_j}\left[\frac{1}{1 + i\tau(\omega_j + \omega_k)} - \frac{1}{1 + i\tau\sum_l \omega_l}\right]. \nonumber
\end{align}
The term $G^*_{3,\mathcal{D}}$ represents the contribution due to nonlinearity in the tube destruction function $\mathcal{D}(v)$:
\begin{align}
    & G^*_{3,\mathcal{D}}(\omega_1,\omega_2,\omega_3) = \\
    & \quad\quad\quad\quad\quad \frac{2 G_0}{45}\sum_j \sum_{k\neq j} \frac{i\tau\omega_j\omega_k}{\omega_j + \omega_k}\frac{1}{1 + i\tau\omega_j}\left[\frac{1}{1 + i\tau(\omega_j + \omega_k)} \right. \nonumber \\
    & \quad\quad\quad\quad\quad\qquad\qquad\qquad \left. - \frac{1}{1 + i\tau\sum_l \omega_l} + \frac{1}{1 + i\tau\omega_{6-j-k}} - 1\right]. \nonumber
\end{align}
The term $G^*_{3,\boldsymbol{Q}}$ represents the contribution due to nonlinearity in the $\boldsymbol{Q}(\boldsymbol{E}_{t't})$ tensor. Because, in simple shear, $\boldsymbol{Q}(\boldsymbol{E}_{t't})$ may be written as a quadratic polynomial in the accumulated strain (see the Supporting Information), this factor arises as a time-strain separable contribution to the third order complex modulus. It may therefore be expressed in terms of the linear modulus (equation \ref{eq:linear}):
\begin{align}
    & G^*_{3,Q}(\omega_1,\omega_2,\omega_3) = \\
    & \quad\quad\quad\quad\quad -\frac{1}{7} \left[G^*\left(\sum_j\omega_j\right) - \sum_j G^*\left(\sum_{k\neq j}\omega_k\right) + \sum_j G^*(\omega_j)\right]. \nonumber
\end{align}
Finally, the parameter $\alpha$ is defined in terms of the limiting slope of $\mathcal{D}(v)$ (or $\mathcal{B}(v)$):
\begin{equation}
    \alpha \equiv \left.\frac{d\mathcal{D}}{dv}\right\rvert_{v = 0} = 1 + \left.\frac{d\mathcal{B}}{dv}\right\rvert_{v = 0},
\end{equation}
where $v = 0$ in simple shear corresponds to the limit of either zero shear-rate or zero shear stress.